\documentclass{article}

\usepackage[preprint]{neurips_2026}
\setcitestyle{numbers,square,comma,sort&compress}

\usepackage{fontspec}
\defaultfontfeatures{Ligatures=TeX}

\usepackage{hyperref}
\usepackage{url}
\usepackage{graphicx}
\usepackage{booktabs}
\usepackage{amsfonts}
\usepackage{amsmath}
\usepackage{nicefrac}
\usepackage{microtype}
\usepackage{xcolor}
\usepackage{algorithm}
\usepackage{algorithmic}
\usepackage{subcaption}
\usepackage{lilyglyphs}
\usepackage[most]{tcolorbox}

\newcommand{\NoteToken}[1]{\texttt{\detokenize{#1}}}

\title{VocalParse: Towards Unified and Scalable Singing Voice Transcription with Large Audio Language Models}

\author{%
  {%
    \begin{tabular}[t]{@{}c@{}}
      \begin{tabular}[t]{@{}c@{\hspace{1.8em}}c@{\hspace{1.8em}}c@{}}
        Yukun Chen\textsuperscript{1,2} &
        Tianrui Wang\textsuperscript{3,2} &
        Zhaoxi Mu\textsuperscript{4,5}
      \end{tabular} \\
      \begin{tabular}[t]{@{}c@{\hspace{2.4em}}c@{}}
        Xinyu Yang\textsuperscript{1}\thanks{Corresponding authors} &
        EngSiong Chng\textsuperscript{2}\footnotemark[1]
      \end{tabular} \\
      {\normalfont \textsuperscript{1}Xi'an Jiaotong University \quad \textsuperscript{2}Nanyang Technological University} \\
      {\normalfont \textsuperscript{3}Tianjin University \quad \textsuperscript{4}Ant Group \quad \textsuperscript{5}Zhejiang University} \\
      {\normalfont \texttt{chenyk@stu.xjtu.edu.cn}}
    \end{tabular}%
  }
}

\begin{document}

\maketitle

\begin{abstract}
High-quality singing annotations are fundamental to modern Singing Voice Synthesis (SVS) systems. However, obtaining these annotations at scale through manual labeling is unrealistic due to the substantial labor and musical expertise required, making automatic annotation highly necessary. Despite their utility, current automatic transcription systems face significant challenges: they often rely on complex multi-stage pipelines, struggle to recover text-note alignments, and exhibit poor generalization to out-of-distribution (OOD) singing data. To alleviate these issues, we present VocalParse, a unified singing voice transcription (SVT) model built upon a Large Audio Language Model (LALM). Specifically, our novel contribution is to introduce an interleaved prompting formulation that jointly models lyrics, melody, and word-note correspondence, yielding a generated sequence that directly maps to a structured musical score. Furthermore, we propose a Chain-of-Thought (CoT) style prompting strategy, which decodes lyrics first as a semantic scaffold, significantly mitigating the context disruption problem while preserving the structural benefits of interleaved generation. Experiments demonstrate that VocalParse achieves state-of-the-art SVT performance on multiple singing datasets. The source code and checkpoint are available at \url{https://github.com/pymaster17/VocalParse}.
\end{abstract}

\section{Introduction}

\begin{figure}[t]
    \centering
    \includegraphics[width=\textwidth]{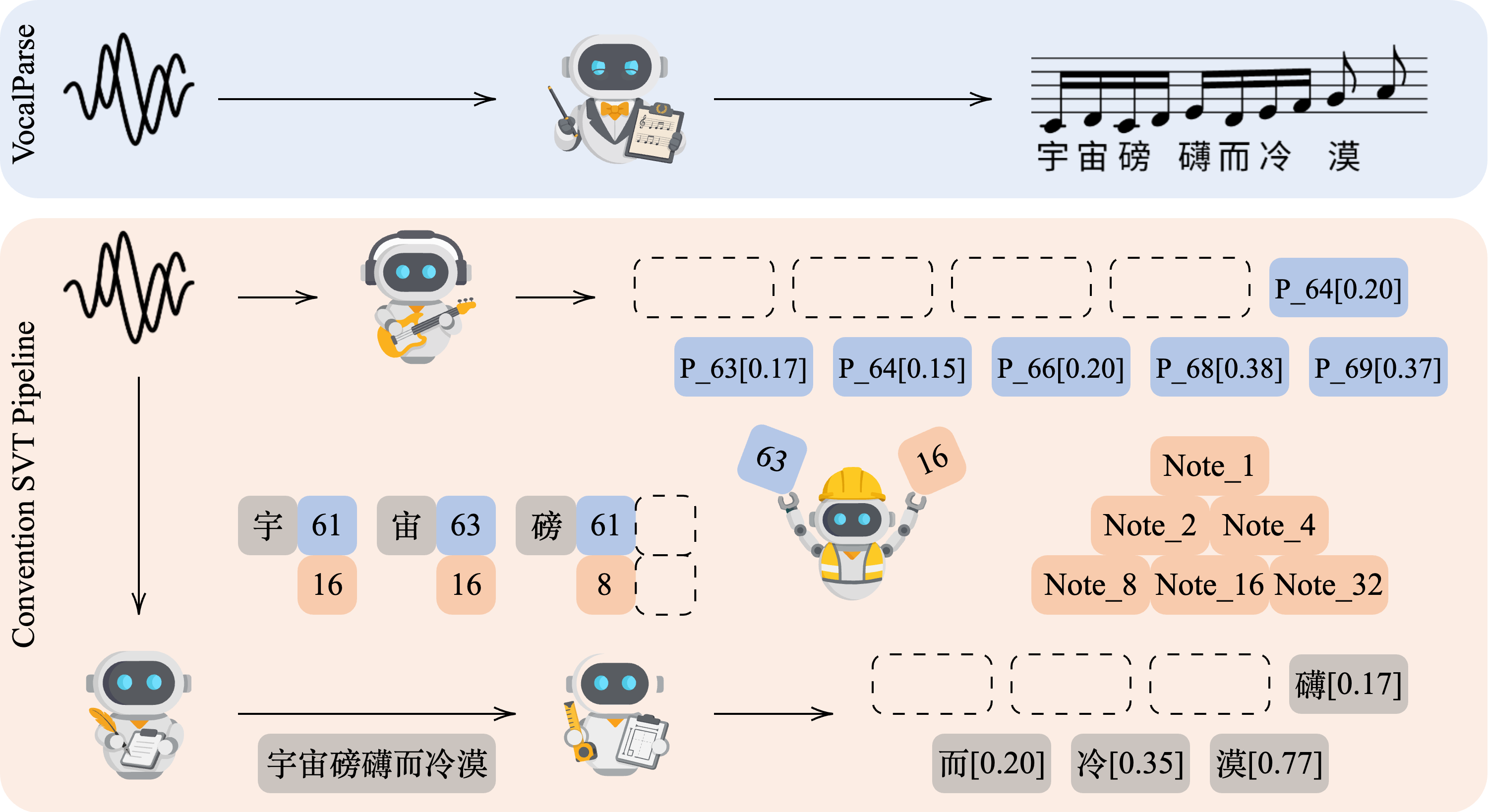}
    \caption{Comparison of VocalParse and conventional SVT pipeline}
    \label{fig:pipeline}
\end{figure}

As SVS systems have evolved from cascaded acoustic-vocoder pipelines~\cite{liu2022diffsinger,zhang2022visinger,guo2025techsinger} to end-to-end transformer-based models~\cite{zhang2025tcsinger,qian2026soulx,zheng2025yingmusic}, their demand for large-scale, well-annotated training data has grown substantially. However, singing annotations remain difficult and expensive to obtain: manual labeling requires both musical expertise and substantial labor, while publicly available singing datasets are still limited in scale~\cite{wang2022opencpop,zhang2022m4singer,zhang2024gtsinger}. This data bottleneck has become a major obstacle to building more controllable and expressive SVS systems.

To reduce annotation cost, prior works often construct automatic labeling pipelines by combining modules such as Automatic Speech Recognition (ASR)~\cite{radford2023robust,gao2022paraformer,an2024funaudiollm}, forced alignment~\cite{mcauliffe2017montreal}, and melody transcription~\cite{wang2022musicyolo,li2024robust}. While practical, such pipelines suffer from three limitations. First, the overall annotation process is decomposed into multiple dependent stages, making the system prone to cascading errors and cumbersome to scale, since manual inspection and correction are often needed to ensure label quality~\cite{wang2022opencpop,zhang2024gtsinger}. Second, lyrics and musical notes are usually predicted separately, so the text-to-note correspondence must be rebuilt through additional alignment procedures~\cite{li2024robust}. Third, many components are adapted from speech or trained on limited singing data~\cite{ou2022transfer,gao2023self}, leading to poor generalization to out-of-distribution (OOD) singing data with large pitch variations, prolonged vowels, and diverse vocal styles. Although recent studies~\cite{li2024robust,guo2025stars,wang2023adapting,wu2024songtrans} have attempted to partially unify the transcription pipeline by integrating some of the above modules into neural models, the fundamental limitations are still not fully resolved. As a result, building a unified, scalable, and robust singing voice transcription system remains an open challenge.

Large Audio Language Models (LALMs) offer a promising foundation for this challenge. Their strong audio-semantic modeling ability makes them attractive for jointly transcribing lyrics and melody within a single autoregressive framework~\cite{chu2024qwen2,ma2024foundation,yan2025ming}. However, existing singing datasets are far smaller than the data typically required to effectively adapt large audio-language models~\cite{shi2024singing,pan2025synthetic}, limiting their robustness and OOD generalization.

To address these challenges, we present VocalParse, a unified and scalable singing voice transcription model built on top of a LALM. First, we introduce SingCrawl, a scalable web-based data pipeline that collects vocal audio and automatically constructs large-scale pseudo labels for singing transcription. Second, we design a structured formulation based on interleaving lyric and music tokens intrinsically reflecting the hierarchical correspondence between words and notes. Third, we propose a Chain-of-Thought (CoT) styled prompting strategy that restores continuous semantic context before structured interleaved decoding, thereby preserving the pretrained ASR capability of the backbone while enabling joint lyric-melody generation. With this design, VocalParse supports both audio-only transcription and lyric-conditioned transcription within the same model, without requiring architectural modifications. Experiments show that VocalParse achieves state-of-the-art performance across lyric, alignment, pitch, and note-related metrics.

Our contributions are three-fold:
\begin{itemize}
    \item We develop \textbf{SingCrawl}, a scalable singing voice crawling, processing and labeling pipeline, constructing a large-scale annotated dataset for Singing Voice Transcription.
    \item We propose \textbf{VocalParse}, a simple and unified SVT model based on LALMs, achieving state-of-the-art performance without complex post-processing or multi-path decoding structures.
    \item We introduce a \textbf{CoT-style prompting} tailored to structured singing transcription with LALMs, which improves compatibility between interleaved generation and pretrained semantic decoding, while naturally enabling optional lyric-conditioned inference.
\end{itemize}

\section{Related Work}

\subsection{Singing Voice Transcription}

Existing singing voice datasets are extremely limited in size~\cite{wang2022opencpop,liu2022diffsinger,huang2021multi} compared with speech~\cite{emilia}, and many of them lack fine-grained annotations. To scale the dataset without relying on heavy labor work, Singing Voice Transcription (SVT) is explored, aiming to automatically extract both semantic information and musical information from vocal recordings. The former corresponds to \textit{Automatic Lyric Transcription} (ALT), which recognizes lyrics, while the latter corresponds to \textit{Automatic Melody Transcription} (AMT), which recovers pitch, note boundaries, and durations. A common practice is to decompose the problem into multiple subtasks, such as lyric transcription, timestamp alignment, and melody transcription, and then combine specialized models into a pipeline. For example, ASR systems such as Whisper~\cite{radford2023robust} and Paraformer~\cite{gao2022paraformer} are often used to obtain initial lyrics, while forced alignment systems such as MFA~\cite{mcauliffe2017montreal} or SOFA provide word boundaries, and melody transcription models such as MusicYOLO~\cite{wang2022musicyolo} or ROSVOT~\cite{li2024robust} generate pitch and note information.

To simplify conventional pipelines, recent work has explored unified singing transcription models. \cite{wang2023adapting} adapt a pretrained speech model with an additional alignment head to jointly predict lyrics and timestamps. SongTrans~\cite{wu2024songtrans} further moves toward joint transcription of lyrics and melody, but relies on a cascaded AR-NAR design with separate modules for different prediction stages. STARS~\cite{guo2025stars} unifies several singing-related predictions in one framework, but still depends on external lyric transcription as an input condition. However, there is still a large gap from existing systems to fully end-to-end transcription. They require external models or additional conditions to work, and built on complex multi-module architectures that are difficult to scale.

\subsection{Large Audio Language Models (LALMs)}

Large Audio Language Models (LALMs) extend text LLMs to audio-based understanding by aligning audio and text representations within a shared modeling framework~\cite{ji2024wavchat}. Depending on the design, this alignment can be achieved either mainly in the audio tokenizer~\cite{alexandre2024moshi,xu2025qwen3} or directly in the language model through interleaved or parallel prompting~\cite{ding2025kimi,wu2025step}. After multimodal adaptation and task-specific finetuning~\cite{chu2024qwen2}, LALMs have achieved strong performance in ASR~\cite{bai2024seed,xu2025fireredasr,shi2026qwen3} and general audio understanding~\cite{ding2025kimi,ghosh2025audio}. Recent studies have further demonstrated their promise in music-related tasks, including song structure analysis~\cite{tan2025songprep,hao2025songformer,mossmusic2026} and music captioning~\cite{ghosh2025music,wang2024muchin}.

Beyond modality fusion, interleaved representations are also attractive for structured music-related tasks~\cite{kim2025note,yuan2025yue}, since they can encode local structural and alignment relations directly in the generated sequence. At the same time, Chain-of-Thought reasoning has begun to be explored in LALMs~\cite{ma2025audio,tian2025step}, suggesting that prompting design can substantially affect how such models use audio context. However, prior LALM work has not addressed unified singing voice transcription with these prompting strategies.

\section{VocalParse}

\begin{figure*}[htp]
    \centering
    \includegraphics[width=\textwidth]{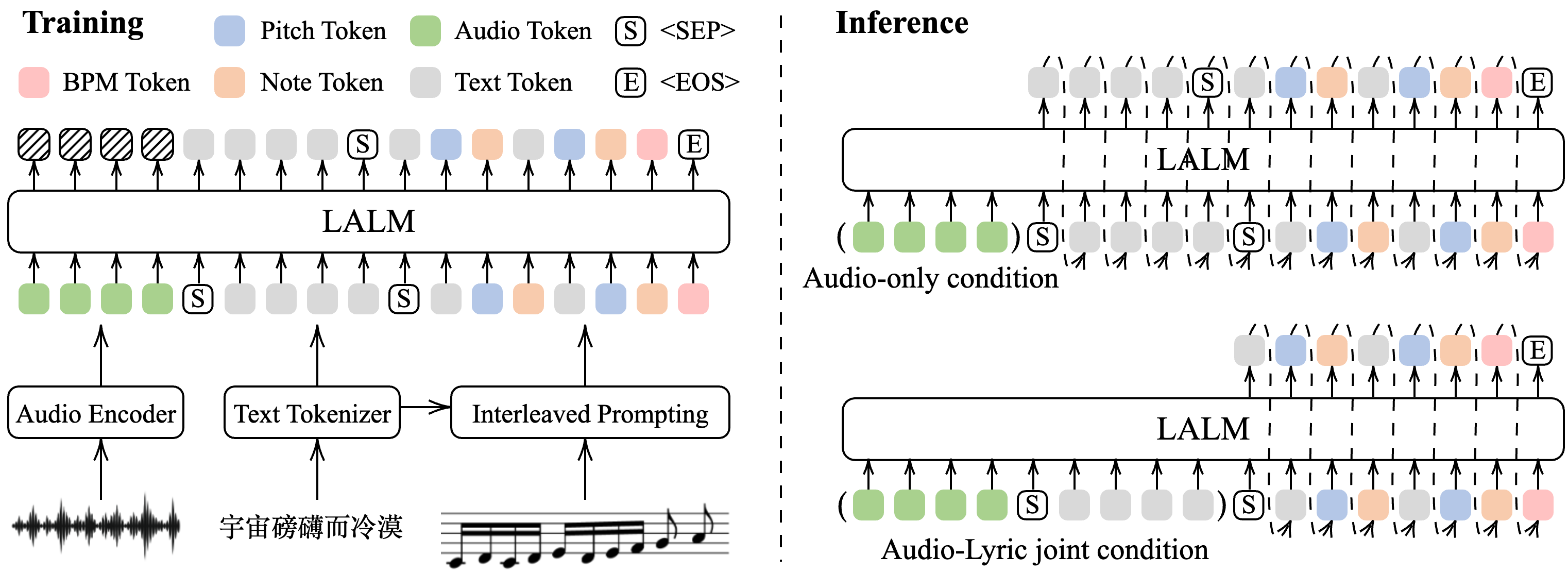}
    \caption{Overview of VocalParse. Left: training paradigm of VocalParse with interleaved word-note supervision and CoT-style prompting. Right: two inference modes of the unified model, including audio-only inference and audio-lyric joint inference.}
    \label{fig:main}
\end{figure*}

\subsection{Overview}
VocalParse reformulates singing voice transcription as a unified autoregressive generation problem over structured symbolic sequences. Given a singing segment, the model aims to transcribe both the lyric content and the corresponding melody, while preserving the word-note correspondence required by downstream singing applications. To this end, we build VocalParse on top of Qwen3-ASR~\cite{shi2026qwen3}, a Large Audio Language Model (LALM) with strong audio-semantic modeling ability inherited from ASR-oriented finetuning, while also retaining basic music understanding from Qwen3-Omni~\cite{xu2025qwen3}.

As shown in Figure~\ref{fig:main}, the input waveform is first converted into discrete audio tokens at 12.5 Hz by the audio tokenizer, and the model is trained to predict a two-stage target sequence. The first stage is a \emph{pure lyric sequence}, which restores a continuous semantic decoding context compatible with the pretrained LALM. The second stage is an \emph{interleaved lyric-note sequence}, which explicitly encodes the local correspondence between each word and its associated notes. In this way, VocalParse jointly models semantic recognition and musical transcription within a single causal decoding framework.

This design is motivated by a key trade-off in unified singing transcription. On the one hand, the output sequence should preserve the natural hierarchical structure of singing, where each word may correspond to one or multiple notes. On the other hand, directly interleaving lyric and music tokens can disrupt the continuous text context that pretrained LALMs rely on for accurate semantic decoding. VocalParse resolves this trade-off through two complementary components: \textbf{interleaved prompting}, which provides a structurally faithful representation of word-note alignment, and \textbf{CoT-style prompting}, which provides a semantic scaffold before structured generation. The two components are introduced next.

\subsection{Interleaved Prompting}

A complete singing transcription should preserve both the lyric content and the corresponding musical realization. In singing voice, these two parts are not independent: each note is associated with a specific word, and each word may span one or multiple notes. This induces a natural hierarchical structure in which lyrics serve as the semantic units and notes describe their local melodic realization. Conventional SVT systems often predict lyrics and melody in separate stages, making it difficult to recover this word-note correspondence in a unified and lossless manner.

To explicitly encode this structure, we introduce an \textbf{interleaved prompting} format, in which each word is immediately followed by its associated note sequence. Let the complete structured transcription be denoted as $\mathcal{S}_{il}$, consisting of $N$ lyric words. For the $i$-th word $w_i$, we attach a corresponding musical sequence $\mathcal{M}_i$, yielding

\begin{equation}
\mathcal{S}_{il} = \bigoplus_{i=1}^{N} \Big[ w_{i} \oplus \mathcal{M}_{i} \Big],
\end{equation}

where $\oplus$ denotes token/sequence concatenation. The musical sequence assigned to $w_i$ contains $K_i$ consecutive notes:
\begin{equation}
\mathcal{M}_i = \bigoplus_{j=1}^{K_i} \left( p_{i,j} \oplus n_{i,j} \right),
\end{equation}
where $p_{i,j}$ and $n_{i,j}$ denote the discrete pitch token and duration token of the $j$-th note aligned to word $w_i$. Here, $K_i = 1$ corresponds to a standard one-to-one word-note mapping, while $K_i > 1$ represents melisma.

This formulation preserves the local structure of singing transcription in a sequence-native manner: each word and its associated notes are placed in close proximity in the decoding stream, allowing the autoregressive model to directly learn the word-note correspondence. To ensure that the generated sequence can be converted into a symbolic score without ambiguity, we define a vocabulary of $128$ \texttt{<PITCH>} tokens corresponding to standard MIDI numbers and $12$ \texttt{<NOTE>} tokens representing note durations from demisemiquaver to semibreve. The detailed definitions of note token are in Appendix Table~\ref{tab:note_token_duration}.

Besides these word-local note tokens, we further introduce a song-level \texttt{<BPM>} token to represent the global tempo. Unlike $(p_{i,j}, n_{i,j})$, which are attached to individual words, the BPM token appears only once in the whole sequence as a suffix.

While this interleaved representation is structurally desirable for unified SVT, it also introduces a new challenge for pretrained LALMs: the continuous text context is interrupted by music tokens. As a result, although interleaving is a natural representation for singing structure, it is not fully compatible with the semantic decoding behavior that ASR-oriented LALMs are pretrained to perform. This motivates the CoT-style prompting strategy described next.

\subsection{CoT-Style Prompting}

\begin{figure}[t]
    \centering
    \includegraphics[width=\textwidth]{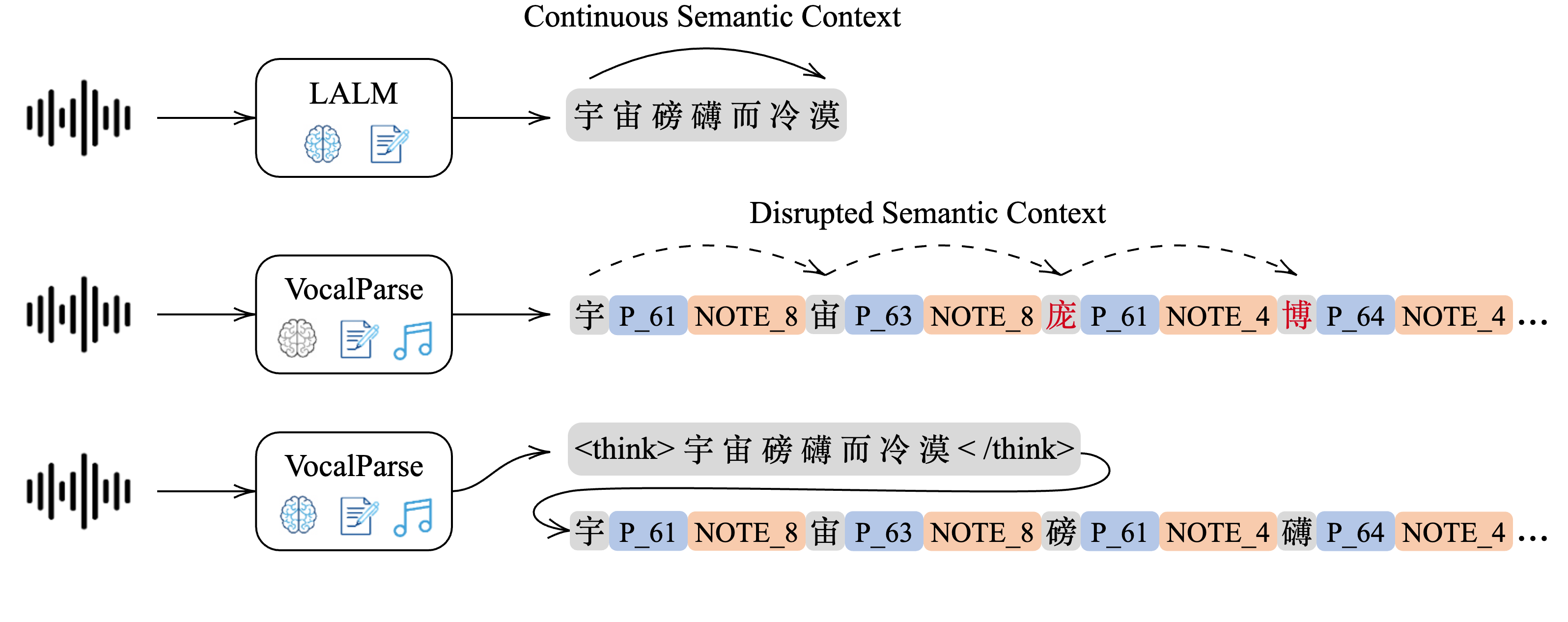}
    \caption{Illustration of CoT-style prompting. Top: standard ASR decoding. Middle: direct interleaved lyric-note decoding. Bottom: CoT-style decoding.}
    \label{fig:cot}
\end{figure}

The advantage of interleaved prompting is that it preserves the fine-grained correspondence between words and notes. However, directly training a pretrained LALM to generate such an interleaved sequence can harm semantic decoding. In a standard lyric sequence $\mathcal{W} = \bigoplus_{i=1}^{N} w_i$, the prediction of the next word mainly depends on preceding lyric tokens and the input audio, i.e.,
\begin{equation}
P(w_i \mid w_{<i}, A).
\end{equation}
By contrast, under direct interleaved decoding, the prediction of $w_i$ is conditioned not only on preceding words but also on intervening music tokens:
\begin{equation}
P(w_i \mid w_{<i}, \mathcal{M}_{<i}, A).
\end{equation}
This changes the local token transition pattern from continuous text to mixed text--music sequences, which is mismatched with the pretraining distribution of ASR-oriented LALMs. In practice, such a mismatch breaks the semantic cue and lead to more homophone errors when model relies on acoustic cue along. In addition, inserting music tokens between adjacent words increases their relative positional distance, which can further weaken the effective semantic dependency across the lyric sequence.

To mitigate this issue, we introduce a \textbf{Chain-of-Thought (CoT) style prompting} strategy that restores semantic continuity before structured decoding. Specifically, we prepend the pure lyric sequence $\mathcal{W}$ as semantic scaffold before the interleaved sequence $\mathcal{S}_{il}$, and form the final target as
\begin{equation}
\mathcal{S}_{cot} = \mathcal{W} \oplus \mathcal{S}_{il}.
\end{equation}
Under this formulation, the generation process can be factorized into two sequential stages:
\begin{equation}
P(\mathcal{S}_{cot} \mid A) = P(\mathcal{W} \mid A)\, P(\mathcal{S}_{il} \mid \mathcal{W}, A).
\end{equation}
In the first stage, the model performs lyric decoding under a purely textual context, which is much more compatible with its pretrained ASR behavior. In the second stage, the model generates the interleaved word-note sequence conditioned on the already decoded global lyric sequence. The pure lyric prefix stabilizes semantic recognition, while the subsequent interleaved sequence enables structured lyric-melody transcription within the same autoregressive framework. The decoding strategies of VocalParse are illustrated in Figure~\ref{fig:cot}.

\subsection{Training and Inference}
Based on the above sequence design, VocalParse is trained with a standard causal language modeling objective. For each sample, the input waveform is first converted into discrete audio tokens, and the target side is constructed as the CoT sequence, consisting of a pure lyric sequence followed by an interleaved word-note sequence. During training, the model is optimized to predict all target tokens in this unified autoregressive stream. More importantly, the CoT formulation makes VocalParse \emph{natively} support two inference modes without any architectural modification or retraining.

\textbf{Audio-only condition.}
When only vocal audio is provided, VocalParse performs full autoregressive decoding from audio tokens to the complete CoT target. The model first generates the pure lyric sequence, acting as an ASR model for singing voice, and then continues to decode the interleaved word-note sequence conditioned on the generated lyrics. In this way, complete singing transcription is accomplished within a single unified decoding process.

\textbf{Audio-Lyric joint condition.}
When reliable lyrics are already available, we directly provide the pure lyric sequence as a prefix and let the model continue decoding only the interleaved word-note sequence. In other words, the first-stage lyric sequence is no longer predicted autoregressively, but explicitly supplied as semantic context. This naturally avoids error propagation from lyric recognition to melody transcription, and allows the model to focus on note prediction conditioned on accurate lyric information, leading to more precise melody transcription.

\section{SingCrawl}

Large-scale data is essential for the robustness and generalization of modern sequence models, yet publicly available singing datasets remain limited in both scale and annotation completeness. This limitation is particularly restrictive for VocalParse, which requires not only lyric supervision but also word-note correspondence. To address this bottleneck, we introduce \textbf{SingCrawl}, a scalable web-based pipeline that converts raw online songs into pseudo-labeled singing-transcription data for VocalParse training. SingCrawl consists of three stages: \textit{Pre-filtering}, \textit{Audio Processing}, and \textit{Automatic Annotation}. The detailed process of SingCrawl is introduced in Appendix~\ref{sec:appendix_singcrawl} and the code can be found at \url{https://github.com/pymaster17/SingCrawl}.

\subsection{Pre-filtering}

The goal of pre-filtering is to retain songs that are suitable for large-scale singing transcription while reducing the cost of downstream processing. We first select candidate songs according to metadata constraints, including language, style tags, and lyric availability. In particular, we exclude tracks that are likely to be instrumental or non-vocal-dominant, and retain only songs with sentence-level lyrics in the metadata, since such information is required for subsequent segmentation and alignment. We further prioritize songs and singers with higher-quality metadata and recording conditions. In this work, we focus on Mandarin songs for efficiency and consistency, although the pipeline itself is not limited to a single language.

\subsection{Audio Processing}

The audio processing stage converts each full song into clean singing segments paired with corresponding lyric excerpts. Starting from full-song audio and sentence-level lyric metadata, we first refine rough segment boundaries using the provided timestamps together with waveform-based silence detection, producing more reliable sentence-level singing excerpts. We then apply vocal extraction and dereverberation to suppress accompaniment and environmental artifacts, so that the resulting audio is more suitable for alignment and note transcription. Finally, a quality control step is applied to remove low-quality segments introduced by source separation or segmentation errors.

\subsection{Automatic Annotation}

The automatic annotation stage generates complete lyric-melody supervision for each processed singing segment. To obtain reliable word-level timestamps, we retrain a singing-oriented forced-alignment model based on SOFA\footnote{\url{https://github.com/qiuqiao/SOFA}} using a mixture of \textit{weak-label} and \textit{full-label} data. Specifically, the crawled web data provides audio together with the corresponding phoneme sequence and is therefore treated as weak-label data, while GTSinger~\cite{zhang2024gtsinger} and M4Singer~\cite{zhang2022m4singer} provide precise phoneme-level timestamp annotations and are used as full-label data. This mixed supervision allows the aligner to benefit from both the scale of weakly labeled web data and the temporal precision of fully annotated singing corpora.

After retraining, the aligner is applied to the processed SingCrawl segments to generate word-level timestamps. Based on these word boundaries, we further use ROSVOT~\cite{li2024robust} to estimate synchronized note boundaries and pitch trajectories for each word. The resulting note sequence is then converted into the discrete symbolic representation used by VocalParse, including pitch tokens, duration tokens, and a song-level BPM token. In this way, SingCrawl produces training targets that are directly compatible with the interleaved word-note formulation of VocalParse.

\section{Experiments}

\subsection{Experimental Setup}
\textbf{Training Details} We train VocalParse on a 2000-hour singing dataset collected through SingCrawl, together with two open-source datasets, GTSinger~\cite{zhang2024gtsinger} (Chinese subset) and M4Singer~\cite{zhang2022m4singer}, which contribute approximately 50 hours in total. We evaluate VocalParse on several widely used singing datasets, including Opencpop~\cite{wang2022opencpop}, ACE-KiSing~\cite{shi2024singing}, OpenSinger~\cite{huang2021multi}, and PopCS~\cite{liu2022diffsinger}. Due to differences in annotation formats across datasets, Opencpop and ACE-KiSing are used for AMT evaluation, while Opencpop, OpenSinger, and PopCS are used for ALT evaluation. Beyond intrinsic SVT benchmarks, we further examine whether the annotations generated by VocalParse can serve as effective supervision for downstream SVS training. Full training details for VocalParse and SVS experiment are provided in Appendix~\ref{sec:appendix_training_details} and~\ref{sec:svs_exp}.

\textbf{Evaluation Metrics} Following standard evaluation protocols, the ALT performance is measured by the Word Error Rate (WER). For AMT task, we evaluate pitch accuracy and temporal accuracy using Mean Absolute Error (MAE). Specifically, $MAE_{pitch}$ is computed as the absolute error in MIDI number, which is already defined in a logarithmic pitch space. For melody evaluation, we report $MAE_{note}$ on note value and $MAE_{dur}$ on nominal duration, following the note definitions in Appendix Table~\ref{tab:note_token_duration}. Formally,

\begin{equation}
\begin{aligned}
MAE_{pitch} &= |MIDI_{pred} - MIDI_{gt}| \\
MAE_{note}  &= \left|\log_{2}\!\left(Note_{pred}^{v}\right) - \log_{2}\!\left(Note_{gt}^{v}\right)\right| \\
MAE_{dur}   &= \left|\log_{2}\!\left(Note_{pred}^{d}\right) - \log_{2}\!\left(Note_{gt}^{d}\right)\right|
\end{aligned}
\end{equation}

where $Note^{v}$ denotes the symbolic note value in quarter-note units, and $Note^{d}$ denotes the BPM-conditioned nominal duration in seconds, i.e., $Note^{d} = \frac{60}{\text{BPM}} \cdot Note^{v}$. Thus, $MAE_{note}$ measures relative melody deviation in symbolic space, while $MAE_{dur}$ measures the resulting absolute-time deviation after normalized by the song-level BPM token. We additionally report $Num_{note}$, the absolute error in the total number of predicted notes per excerpt, to measure structural segmentation accuracy.

\textbf{Baselines.} We compare VocalParse against multiple representative baselines. For AMT, we benchmark against ROSVOT~\cite{li2024robust}, MusicYOLO~\cite{wang2022musicyolo}, and STARS~\cite{guo2025stars}. Since these systems require additional conditions beyond audio, we provide ground-truth lyrics to all three models, and additionally provide SOFA-predicted timestamps to ROSVOT and MusicYOLO following their common usage. Therefore, the audio-lyric setting of VocalParse serves as the fairest matched-condition comparison against prior AMT systems. For ALT, we compare against LyricWhiz~\cite{zhuo2023lyricwhiz}, a Whisper-based singing transcription model~\cite{wang2023adapting}, and our foundation Qwen3-ASR model~\cite{shi2026qwen3}.

\subsection{Main Results}

\begin{table}[!t]
\centering
\caption{AMT performance on Opencpop and ACE-KiSing.}
\label{tab:amt}
\setlength{\tabcolsep}{1pt}
\begin{tabular*}{\textwidth}{@{\extracolsep{\fill}}l|cccc|cccc@{}}
\toprule
 & \multicolumn{4}{c|}{Opencpop} & \multicolumn{4}{c}{ACE-KiSing\textsuperscript{\dag}} \\
\cmidrule{2-9}
 & {\small $MAE_{pitch}$} & {\small $MAE_{note}$} & {\small $MAE_{dur}$} & {\small $Num_{note}$} & {\small $MAE_{pitch}$} & {\small $MAE_{note}$} & {\small $MAE_{dur}$} & {\small $Num_{note}$} \\
\midrule
STARS & 1.12 & 0.57 & 0.47 & 0.17 & 1.42 & 0.60 & 0.57 & 0.35 \\
MusicYOLO & 0.64 & 0.54 & 0.56 & 0.41 & 1.63 & 0.73 & 0.63 & 0.55 \\
ROSVOT & 0.38 & 0.45 & 0.40 & 0.20 & 1.08 & 0.62 & 0.54 & \textbf{0.23} \\
\midrule
\shortstack[l]{VocalParse\\(Audio-only)} & 0.56 & 0.44 & 0.34 & \textbf{0.11} & \textbf{0.53} & \textbf{0.52} & \textbf{0.49} & 0.29 \\
\shortstack[l]{VocalParse\\(Audio-Lyric)} & \textbf{0.35} & \textbf{0.43} & \textbf{0.33} & \textbf{0.11} & - & - & - & - \\
\bottomrule
\end{tabular*}

\vspace{2pt}
\begin{flushleft}
\footnotesize
\textsuperscript{\dag} The audio-lyric setting is not reported on ACE-KiSing because this dataset only provides phoneme-level annotations.
\end{flushleft}
\end{table}

\begin{table}[!t]
\centering
\caption{ALT performance in WER (\%, lower is better).}
\label{tab:alt}
\begin{tabular}{c|ccc}
\toprule
 & Opencpop & OpenSinger & PopCS \\
\midrule
Qwen3-ASR & \textbf{3.41} & 5.93 & \textbf{7.83} \\
LyricWhiz & 9.68 & 12.76 & 11.64 \\
Whisper-adapted & 8.67 & 16.55 & 21.68 \\
\midrule
VocalParse & 3.79 & \textbf{5.69} & 8.16 \\
\bottomrule
\end{tabular}
\end{table}

\textbf{Automatic Melody Transcription.} The AMT results on Opencpop and ACE-KiSing are shown in Table~\ref{tab:amt}. Under the audio-lyric condition, VocalParse achieves the fairest comparison with prior systems, since the baselines also rely on lyric-related side information. In this matched-condition setting, VocalParse achieves state-of-the-art performance on Opencpop across all reported metrics, reducing $MAE_{pitch}$ to 0.35, $MAE_{note}$ to 0.43, $MAE_{dur}$ to 0.33, and $Num_{note}$ to 0.11. These results indicate that VocalParse can effectively leverage lyric context to improve fine-grained note prediction and word-note correspondence. Notably, VocalParse also surpasses the SOFA+ROSVOT pipeline that serves as the pseudo-label annotator during data construction. This result suggests that VocalParse is not merely imitating the teacher pipeline, but can distill and smooth noisy pseudo labels, producing more stable predictions.

We additionally report an audio-only setting to demonstrate the unified transcription ability of VocalParse. Unlike prior AMT systems that depend on auxiliary textual conditions, VocalParse can directly transcribe both lyrics and melody from audio alone within a single autoregressive framework. Despite using less input information, the audio-only setting remains highly competitive and still outperforms most baselines on structural metrics. On Opencpop, it achieves $MAE_{pitch}=0.56$, $MAE_{note}=0.44$, $MAE_{dur}=0.34$, and $Num_{note}=0.11$, outperforming STARS and MusicYOLO on most metrics and approaching the performance of ROSVOT. On ACE-KiSing, VocalParse also shows strong robustness, outperforming STARS and MusicYOLO across pitch, note, and duration MAE.

\textbf{Automatic Lyric Transcription.} The ALT results are summarized in Table~\ref{tab:alt}. Although VocalParse is trained for unified lyric-and-melody transcription, it preserves strong lyric recognition performance, achieving WERs of $3.79\%$, $5.69\%$, and $8.16\%$ on Opencpop, OpenSinger, and PopCS, respectively. Compared with dedicated singing transcription systems such as LyricWhiz and Whisper-adapted, VocalParse substantially reduces transcription errors across all three benchmarks. Moreover, its ALT performance remains competitive with the ASR-specialized Qwen3-ASR model, indicating that introducing melody modeling does not noticeably compromise lyric recognition ability.

\subsection{Ablation Study}

\begin{table*}[!t]
\centering
\caption{Ablation study on Opencpop. We report both lyric transcription accuracy using WER (\%) and melody transcription quality using pitch, note-value, duration, and note-count errors.}
\label{tab:ablation}
\begin{tabular}{l|ccccc}
\toprule
 & WER (\%) & $MAE_{pitch}$ & $MAE_{note}$ & $MAE_{dur}$ & $Num_{note}$ \\
\midrule
VocalParse & \textbf{3.79} & \textbf{0.56} & \textbf{0.44} & \textbf{0.34} & \textbf{0.11} \\
\quad - w/o CoT & 7.18 & 0.92 & 0.46 & 0.38 & 0.12 \\
\quad - w/o SingCrawl & 4.86 & 0.94 & 0.47 & 0.37 & \textbf{0.11} \\
\bottomrule
\end{tabular}
\end{table*}

To analyze the contributions of our major design choices, we conduct ablation studies on Opencpop, as shown in Table~\ref{tab:ablation}.

\textbf{Effect of CoT-style Prompting.} Removing the CoT prompting strategy (\textit{w/o CoT}) leads to a clear degradation in both lyric transcription and melody prediction. In particular, WER increases substantially from $3.79\%$ to $7.18\%$, while $MAE_{pitch}$ and $MAE_{dur}$ also degrade moderately. These results support our hypothesis that prepending a pure lyric sequence helps preserve the semantic decoding behavior of the pretrained LALM, while still benefiting the subsequent interleaved lyric-note generation.

\textbf{Effect of SingCrawl Pipeline.} Training without large-scale automated curated data via SingCrawl pipeline (\textit{- w/o SingCrawl}) leads to an expected jump in the pitch error ($MAE_{pitch}$ leaps to 0.94) and overall WER (4.86\%). The constrained diversity of manually curated academic datasets significantly limits the model's capacity to capture complex phonetic-melodic intersections and generalized acoustic variance, thus reinforcing the indispensable value of the high-quality, large-scale synthetic alignment data curated via SingCrawl.

\section{Limitations}

Our current BPM estimation in Algorithm~\ref{alg:bpm} assumes a single global tempo for each song segment. This may introduce bias for performances with rubato, ritardando, or natural tempo drift, where a fixed BPM cannot fully capture local timing variation.

The current autoregressive decoding design does not enforce exact consistency between the pure lyric prefix and the lyric tokens in the later interleaved sequence. As a result, the final structured output can drift from correct semantic scaffold in rare cases.

Although VocalParse can distill and smooth noisy pseudo labels and even outperform the teacher pipeline in downstream evaluation, its performance upper bound is still constrained by teacher quality.

Finally, due to computational budget and time constraints, our experiments were conducted exclusively on Mandarin data. While the framework is theoretically generalizable, adapting it to other languages may require additional structural refinement.

\section{Conclusion}

In this paper, we presented \textit{VocalParse}, a unified and scalable singing voice transcription framework built on a Large Audio Language Model and \textit{SingCrawl} to support scalable training. VocalParse formulates singing voice transcription as structured autoregressive generation over interleaved lyric-note sequences, enabling joint modeling of lyrics, melody, and fine-grained word-note correspondence within a single model. To address the semantic disruption introduced by direct interleaving, we further proposed a CoT-style prompting strategy that first establishes continuous lyric context and then performs structured lyric-note decoding. This design preserves the structural advantages of interleaved representation while maintaining strong semantic decoding ability, and naturally supports both audio-only and lyric-conditioned inference.

Overall, our results suggest that LALMs have acquired strong audio understanding capabilities through large-scale pretraining, and, when paired with properly designed adaptation strategies, offer substantial potential for downstream MIR tasks such as SVT.

\bibliographystyle{plainnat}
\bibliography{neurips_2026}

\appendix

\section{Implementation Details of SingCrawl}
\label{sec:appendix_singcrawl}

\begin{figure*}[htp]
    \centering
    \includegraphics[width=\textwidth]{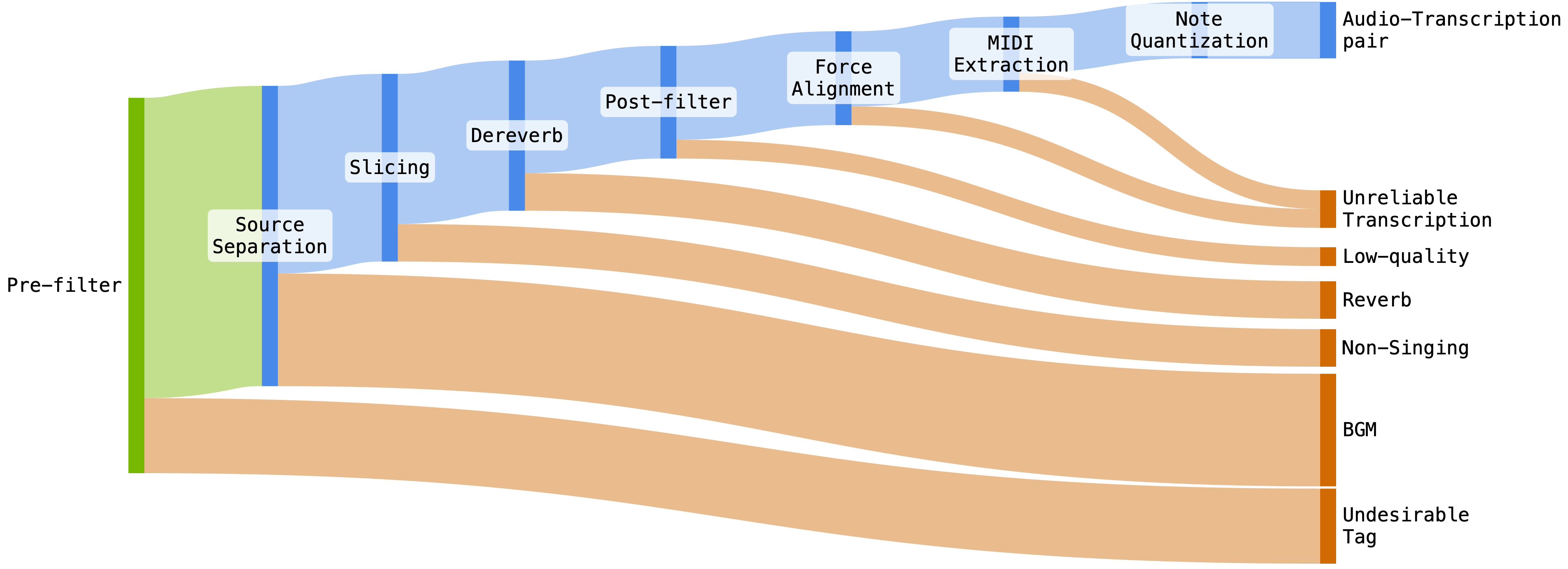}
    \caption{End-to-end data flow of SingCrawl, from raw web songs to the final pseudo-labeled singing segments used for VocalParse training.}
    \label{fig:sankey}
\end{figure*}

As illustrated in Figure~\ref{fig:sankey}, SingCrawl converts raw web songs into paired singing-transcription training data through metadata filtering, audio processing, and automatic annotation.

\paragraph{Metadata filtering.}
Before downloading and processing audio, we first apply a strict metadata-based filter to reduce low-quality or unsuitable samples. We retain only Mandarin solo songs with available sentence-level lyrics, and discard tracks that are likely to be instrumental, chorus-dominant, or non-vocal-centered. To further improve source quality, we remove songs with fewer than 100 likes, singers with fewer than 1{,}000 followers, and audio files below 320 kbps MP3 quality. We also exclude music styles such as rap and rock, which often cause difficulties for downstream alignment and melody extraction tools.

\paragraph{Lyric cleaning and segment slicing.}
The raw crawled data consists of full-song audio together with sentence-level lyric files. We first clean the lyric metadata using a rule-based text filter to remove irrelevant content such as singer names, publisher information, or other non-lyric text. Sentence-level timestamps are then used as initial segmentation anchors. Since these timestamps are often noisy, we refine them by snapping each boundary to its nearest silence region detected from the waveform, producing more reliable sentence-level singing segments.

\paragraph{Vocal extraction and quality control.}
Each segmented song is processed by two mel-RoFormer-based models, \texttt{big\_beta6x.ckpt} for vocal separation and \texttt{dereverb\_mel\_band\_roformer\_anvuew\_sdr\_19.1729.ckpt} for dereverberation, both from MSST-WebUI\footnote{\url{https://github.com/SUC-DriverOld/MSST-WebUI}}. These steps generate relatively dry vocal tracks that are more suitable for alignment and note transcription. We found that they are also the most time-consuming stages in the pipeline, with a real-time factor of roughly 0.1 on an NVIDIA A100 GPU. We additionally experimented with harmony separation, but observed that it often introduced audible distortion to the lead vocal, and therefore did not adopt it in the final pipeline.

\paragraph{Forced alignment.}
Automatic annotation starts from word-level alignment. We retrain SOFA on singing data, using \texttt{PypinyinG2P} instead of the original fixed lexicon to improve scalability to web-scale Mandarin data. We also replace the original UNet block in SOFA with a ConvNeXt-style architecture for more stable large-scale training. During inference, samples with alignment confidence below 0.35 are discarded, since they are typically associated with severe pronunciation ambiguity, strong vocal effects, or other failure cases.

\paragraph{Note extraction and post-processing.}
The word boundaries predicted by the singing-adapted aligner are then provided to ROSVOT as timing conditions to estimate synchronized note boundaries and pitch trajectories. We further apply lightweight post-processing to remove extremely short note regions and merge adjacent notes with the same pitch, reducing fragmentation artifacts introduced by automatic transcription.

\paragraph{Tempo estimation and note quantization.}
After obtaining note durations in seconds, we estimate a song-level BPM and quantize each note into the discrete duration vocabulary used by VocalParse. This produces symbolic note values that are directly compatible with the interleaved lyric--note target sequence. The detailed BPM estimation and duration quantization algorithm is described in Appendix~\ref{sec:note_quantization}.

\paragraph{Data scale.}
In total, we crawled approximately 65k raw songs, corresponding to about 5k hours of audio. After the full processing and filtering pipeline, the resulting dataset contains about 1.7 million singing segments and roughly 2k hours of usable training data.

\section{Note Quantization}
\label{sec:note_quantization}

To support structured lyric-melody generation, VocalParse uses a discrete note vocabulary to represent symbolic duration values. Table~\ref{tab:note_token_duration} summarizes the note tokens used in VocalParse together with their corresponding note values. Here, ``Note Value'' corresponds to $Note^{v}$ in quarter-note units.

\begin{table}[h]
\centering
\caption{Note token definitions used for duration initialization. ``Note Value'' corresponds to $Note^{v}$.}
\label{tab:note_token_duration}
\small
\begin{tabular}{llc}
\toprule
\textbf{Note token} & \begin{tabular}[c]{@{}l@{}}\textbf{Note}\\ \textbf{Value}\end{tabular} & \textbf{Icon} \\
\midrule
\NoteToken{<NOTE_32>} & 0.125 & \thirtysecondNote \\
\NoteToken{<NOTE_DOT_32>} & 0.1875 & \thirtysecondNoteDotted \\
\NoteToken{<NOTE_16>} & 0.25 & \sixteenthNote \\
\NoteToken{<NOTE_DOT_16>} & 0.375 & \sixteenthNoteDotted \\
\NoteToken{<NOTE_8>} & 0.5 & \eighthNote \\
\NoteToken{<NOTE_DOT_8>} & 0.75 & \eighthNoteDotted \\
\NoteToken{<NOTE_4>} & 1.0 & \quarterNote \\
\NoteToken{<NOTE_DOT_4>} & 1.5 & \quarterNoteDotted \\
\NoteToken{<NOTE_2>} & 2.0 & \halfNote \\
\NoteToken{<NOTE_DOT_2>} & 3.0 & \halfNoteDotted \\
\NoteToken{<NOTE_1>} & 4.0 & \wholeNote \\
\NoteToken{<NOTE_DOT_1>} & 6.0 & \wholeNoteDotted \\
\bottomrule
\end{tabular}
\end{table}

To convert automatically predicted note durations into the discrete symbolic targets used by VocalParse, a song-level BPM is required to map absolute time (in seconds) into note values. Since crawled web data rarely provides reliable tempo metadata, we estimate BPM automatically using an EM-like iterative algorithm.

The key assumption is that each observed duration $d_i$ approximately follows
\begin{equation}
d_i \approx k_i \cdot T,
\end{equation}
where $T = 60/\text{BPM}$ is the quarter-note duration and $k_i \in \mathcal{K}$ is one of the standard note multipliers listed in Table~\ref{tab:note_token_duration}. Intuitively, the algorithm searches for the BPM under which the observed note durations can be best explained by a small set of standard symbolic note values.

\begin{figure}[t]
    \centering
    \includegraphics[width=0.7\textwidth]{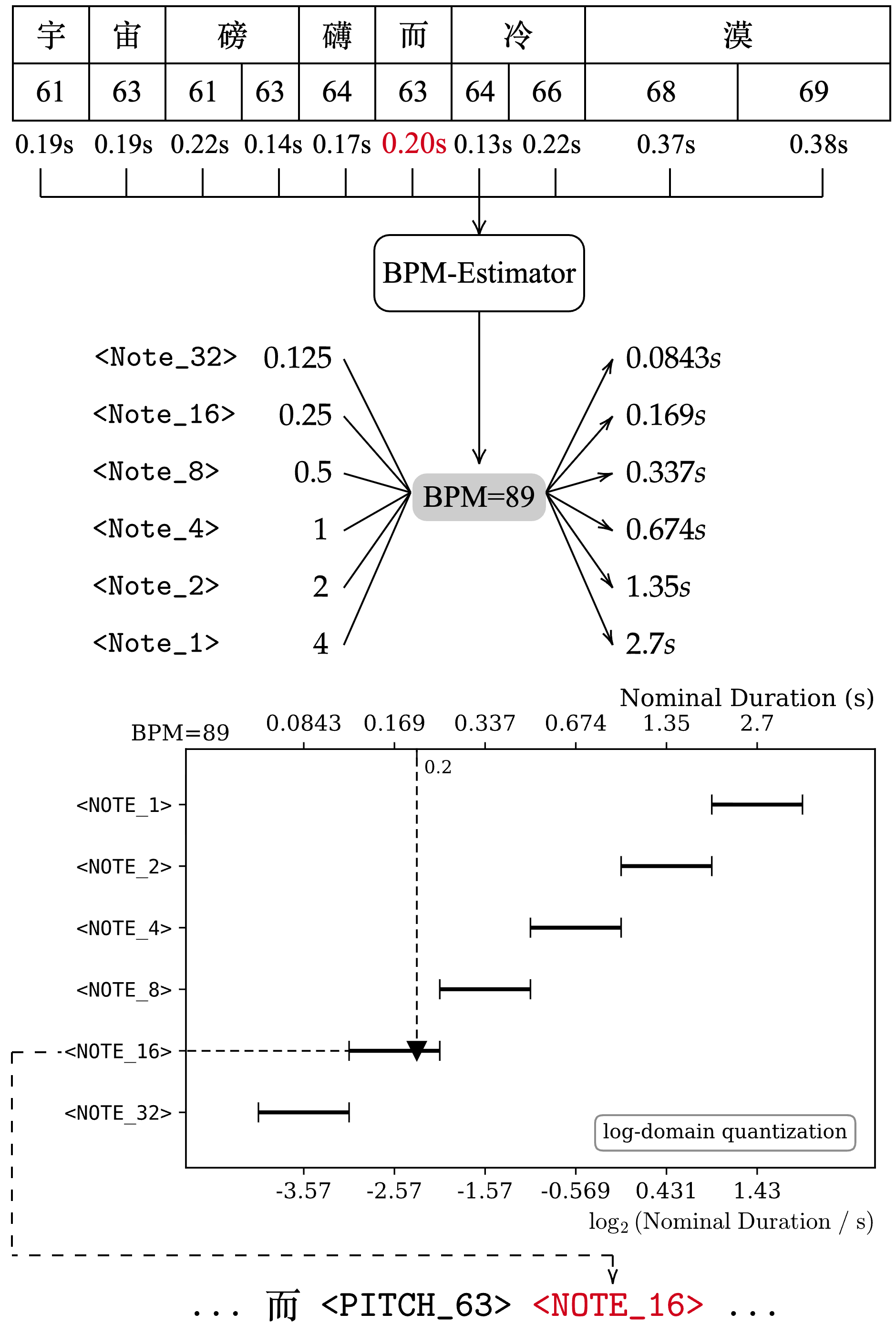}
    \caption{Illustration of the note quantization process, including BPM estimation and mapping from continuous note durations to discrete symbolic note tokens.}
    \label{fig:quantization}
\end{figure}

The procedure starts by building a duration histogram and initializing $T$ from its dominant mode. Since the dominant duration may correspond to different metrical levels, we consider three hypotheses in which the mode corresponds to a quarter note, an eighth note, or a half note. For each hypothesis, we alternate between assigning the closest note multiplier to each duration (E-step) and updating $T$ by least squares (M-step):
\begin{equation}
T \leftarrow \frac{\sum_i d_i \cdot k_i}{\sum_i k_i^2}.
\end{equation}
We then select the hypothesis with the lowest total quantization error,
\begin{equation}
\sum_i (d_i - k_i T)^2,
\end{equation}
and normalize the resulting BPM to the range $[60, 190]$ by octave-equivalent doubling or halving. The complete procedure is given in Algorithm~\ref{alg:bpm}. Figure~\ref{fig:quantization} illustrates the overall quantization process.

\begin{algorithm}[h]
\caption{EM-like BPM Estimation}
\label{alg:bpm}
\begin{algorithmic}[1]
\REQUIRE Note duration sequence $D = \{d_1, \dots, d_N\}$ (seconds)
\ENSURE Estimated BPM

\STATE Filter $D$: retain $0.05\text{s} \le d_i \le 3.0\text{s}$
\STATE Build histogram (bin width $= 0.03$\,s)
\STATE $T_{\text{mode}} \leftarrow$ center of highest-count bin
\STATE $\mathcal{H} \leftarrow \{T_{\text{mode}},\; 2 \cdot T_{\text{mode}},\; T_{\text{mode}} / 2\}$

\FOR{each $T_{\text{init}} \in \mathcal{H}$}
    \STATE $T \leftarrow T_{\text{init}}$
    \FOR{$\text{iter} = 1$ \TO $10$}
        \STATE \textbf{E-step:} $\forall\, i:\; k_i \leftarrow \arg\min_{k \in \mathcal{K}} |d_i / T - k|$
        \STATE \textbf{M-step:} $T_{\text{new}} \leftarrow \frac{\sum_i d_i \cdot k_i}{\sum_i k_i^2}$
        \IF{$|T_{\text{new}} - T| < 0.001$}
            \STATE $T \leftarrow T_{\text{new}}$;\; \textbf{break}
        \ENDIF
        \STATE $T \leftarrow T_{\text{new}}$
    \ENDFOR
    \STATE $\text{err}(T_{\text{init}}) \leftarrow \sum_i (d_i - k_i \cdot T)^2$
\ENDFOR

\STATE $T^* \leftarrow \arg\min_{T_{\text{init}} \in \mathcal{H}} \text{err}(T_{\text{init}})$
\STATE $\text{BPM} \leftarrow 60 / T^*$
\WHILE{$\text{BPM} < 60$}
    \STATE $\text{BPM} \leftarrow \text{BPM} \times 2$
\ENDWHILE
\WHILE{$\text{BPM} > 190$}
    \STATE $\text{BPM} \leftarrow \text{BPM} / 2$
\ENDWHILE

\RETURN $\text{round}(\text{BPM})$
\end{algorithmic}
\end{algorithm}

\section{Training Details of VocalParse}
\label{sec:appendix_training_details}

VocalParse is initialized from the 1.7B-parameter Qwen3-ASR pretrained checkpoint and trained with \emph{full finetuning}. We train the model on 2 NVIDIA H100 GPUs using Distributed Data Parallel (DDP). Training runs for 120k steps and takes approximately 17 hours in total.

To improve hardware utilization under variable-length audio--text pairs, we adopt dynamic batching. The maximum number of batch tokens is set to 18{,}000 per GPU, and each batch contains at most 64 samples on each GPU.

We use a cosine learning-rate schedule with 12k warmup steps. The peak learning rate is set to $2\times10^{-5}$. Unless otherwise specified, all reported results in the main paper use this training configuration.

\section{SVS Experiment}
\label{sec:svs_exp}
To validate the practical value of VocalParse for data annotation, we conduct downstream SVS experiments under different dataset constructions. Since our annotation format is interleaved lyric-note sequences, we adopt an LM-based SVS backbone and choose DiTAR as the baseline architecture due to its strong performance and relatively simple training pipeline.

We consider four training settings. $Ac_1$ uses only academic datasets with ground-truth lyrics and melody labels, namely the Chinese subset of GTSinger and M4Singer, with a total duration of approximately 50 hours. $Scale_M$ and $Scale_L$ represent 200-hour and 2000-hour subsets of SingCrawl respectively, automatically annotated by VocalParse. $Ac_2$ uses OpenSinger recordings and lyrics, while the melody labels are generated by VocalParse. We use OpenSinger rather than GTSinger/M4Singer for this setting because the latter two datasets are already involved in VocalParse training. All four settings use the same SVS model architecture, while the training hyperparameters are adjusted according to dataset scale. For a controlled evaluation, all models are tested on the same Opencpop test set.

We evaluate the generated singing from three perspectives: Aesthetics Quality, Rhythm Similarity, and Melody Similarity. For aesthetics, we report SingMOS~\cite{tang2025singmos} together with CE and PQ from Aesthetics AudioBox~\cite{tjandra2025meta}, and additionally include an AB preference test as a subjective supplement. For rhythm similarity, we compute Boundary Error Rate (BER) and Intersection over Union (IOU) between word alignments extracted from the reference and generated audio. For melody similarity, we report Raw Pitch Accuracy (RPA) between note transcriptions. To reduce teacher-evaluator bias, all alignments and note transcriptions used in evaluation are extracted by STARS~\cite{guo2025stars} rather than the SOFA+ROSVOT pipeline used in data annotation.

\begin{table}[htp]
\centering
\caption{SVS Experiment Results}
\label{tab:svs_experiment}
\begin{tabular}{lcccccc}
\toprule
 & \multicolumn{3}{c}{\textbf{Aesthetics Quality}} & \multicolumn{2}{c}{\textbf{Rhythm SIM}} & \textbf{Melody SIM} \\
\cmidrule(lr){2-4} \cmidrule(lr){5-6} \cmidrule(lr){7-7}
 & \textbf{SingMOS} $\uparrow$ & \textbf{CE} $\uparrow$ & \textbf{PQ} $\uparrow$ & \textbf{BER} $\downarrow$ & \textbf{IOU} $\uparrow$ & \textbf{RPA} $\uparrow$ \\
\midrule
$Ac_1$ & 4.56 & 5.94 & 7.50 & 0.50 & 0.46 & 0.39 \\
\quad $+Scale_M$ & 4.54 & 5.91 & 7.62 & 0.47 & 0.58 & 0.72 \\
\quad $+Scale_L$ & 4.52 & 5.90 & 7.63 & 0.47 & 0.59 & 0.74 \\
\midrule
$Ac_2$ & 4.37 & 6.02 & 7.53 & 0.50 & 0.40 & 0.39 \\
\bottomrule
\end{tabular}
\end{table}

\begin{figure}[t]
    \centering
    \begin{subfigure}[t]{0.49\textwidth}
        \centering
        \includegraphics[width=\linewidth]{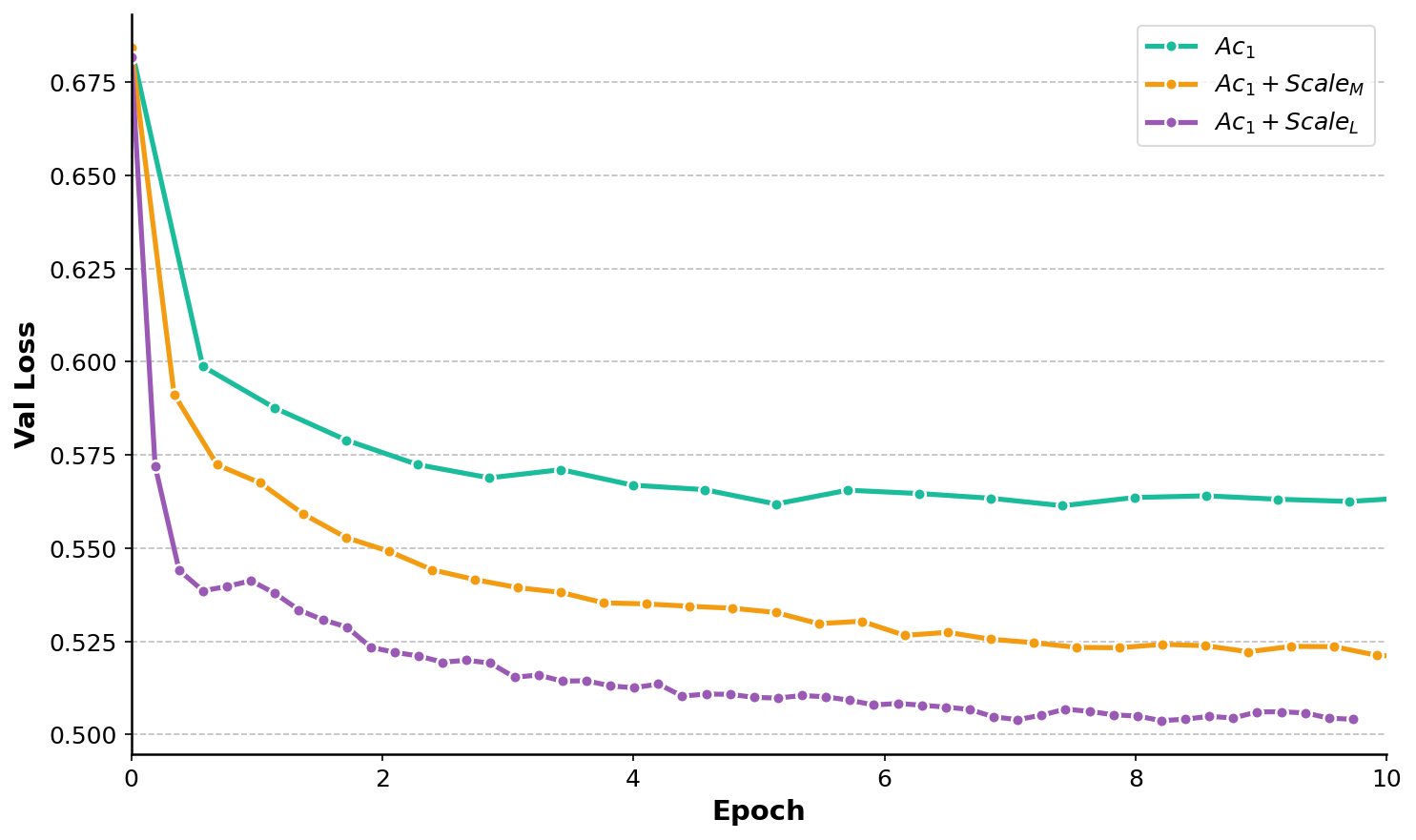}
        \caption{Validation loss of SVS}
        \label{fig:val_loss}
    \end{subfigure}
    \hfill
    \begin{subfigure}[t]{0.49\textwidth}
        \centering
        \includegraphics[width=\linewidth]{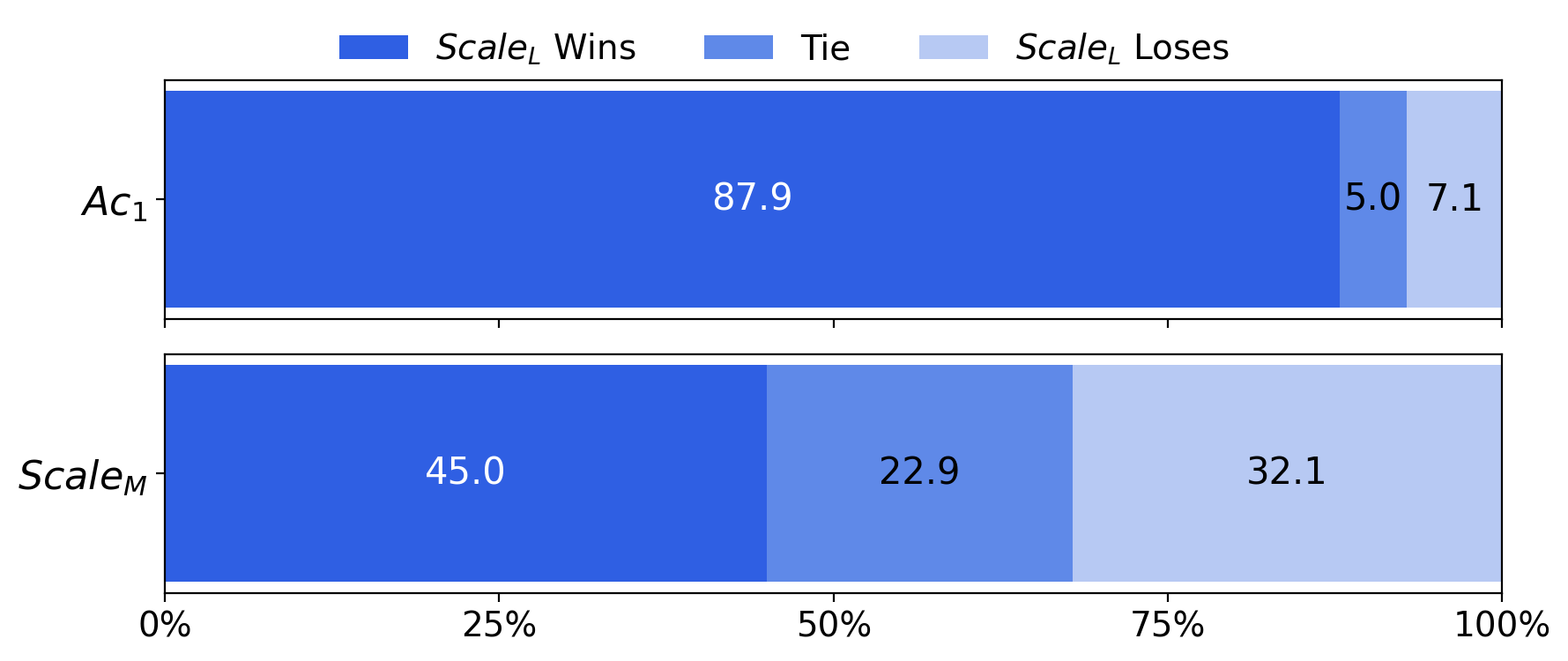}
        \caption{AB Test on different data constructions}
        \label{fig:AB_Test}
    \end{subfigure}
    \caption{SVS results under different data construction settings.}
    \label{fig:svs_results}
\end{figure}

\textbf{Conclusion of SVS Experiments.} The results in Table~\ref{tab:svs_experiment} and Figure~\ref{fig:val_loss} verify the practical value of VocalParse for scalable SVS data annotation. As the amount of automatically labeled SingCrawl data increases from 0 to 200 hours and then to 2000 hours, the validation loss decreases monotonically, suggesting that larger and more diverse pseudo-labeled singing data improves the generalization ability of the downstream SVS model.

More importantly, adding pseudo-labeled data leads to substantial gains in rhythm and melody similarity. Compared with $Ac_1$, $Scale_M$ and $Scale_L$ reduce BER from 0.50 to 0.47/0.47, improve IOU from 0.46 to 0.58/0.59, and boost RPA from 0.39 to 0.72/0.74. These results indicate that large-scale pseudo-labeled data significantly improves the model's ability to follow lyric and melody conditions, while preserving strong synthesis quality.

Although SingCrawl data may contain quality fluctuations due to web crawling and source separation, the aesthetics-related metrics remain largely stable across $Ac_1$, $Scale_M$ and $Scale_L$, with only marginal changes in SingMOS, CE, and PQ. This suggests that the automatically annotated data improves controllability-related metrics without noticeably harming perceptual quality.

Comparing $Ac_2$ and $Ac_1$, which are similar in scale but differ in label source, we observe comparable overall performance, indicating that pseudo labels generated by VocalParse can serve as usable supervision for SVS training even without ground-truth melody annotations. Finally, the AB test in Figure~\ref{fig:AB_Test} provides additional subjective support for the effectiveness of data scaling, with $Scale_L$ being preferred over both $Ac_1$ and $Scale_M$.

\section{Ethics and Responsibility}
\label{sec:ethics}

This work involves large-scale singing data collection and automatic music annotation. We recognize that music recordings, lyrics, and related metadata may be protected by copyright and other legal rights. Therefore, we will not release any concrete crawled data, including raw audio, separated vocals, lyrics, metadata, URLs, or pseudo labels associated with specific songs. All crawled materials are used only for internal research purposes to develop and evaluate the proposed annotation pipeline.

To support open science while respecting music copyright, we release the pretrained weights of VocalParse and the data processing workflow of SingCrawl, including the filtering, segmentation, vocal processing, alignment, and symbolic conversion procedures. The released pipeline will not contain copyrighted music content or song-level identifiers. Researchers who use the pipeline are responsible for ensuring that any data they process is legally obtained and used in accordance with applicable laws, licenses, and platform terms.

\begin{tcolorbox}[
    title=Instructions Shown to Participants,
    colback=gray!5,
    colframe=gray!60,
    fonttitle=\bfseries,
    breakable
]
\textbf{Instructions:}
\begin{itemize}
    \item Most important: use high-quality studio headphones and a good soundcard.
    \item For each test, you will hear two audio samples, A and B, generated from the same input.
    \item Listen to both samples carefully, then select the one you prefer.
    \item If you truly cannot tell any difference, you may select ``Equal''.
    \item Try to rate the overall impression and do not concentrate on single aspects.
\end{itemize}
\end{tcolorbox}

\begin{figure}[htp]
    \centering
    \includegraphics[width=\textwidth]{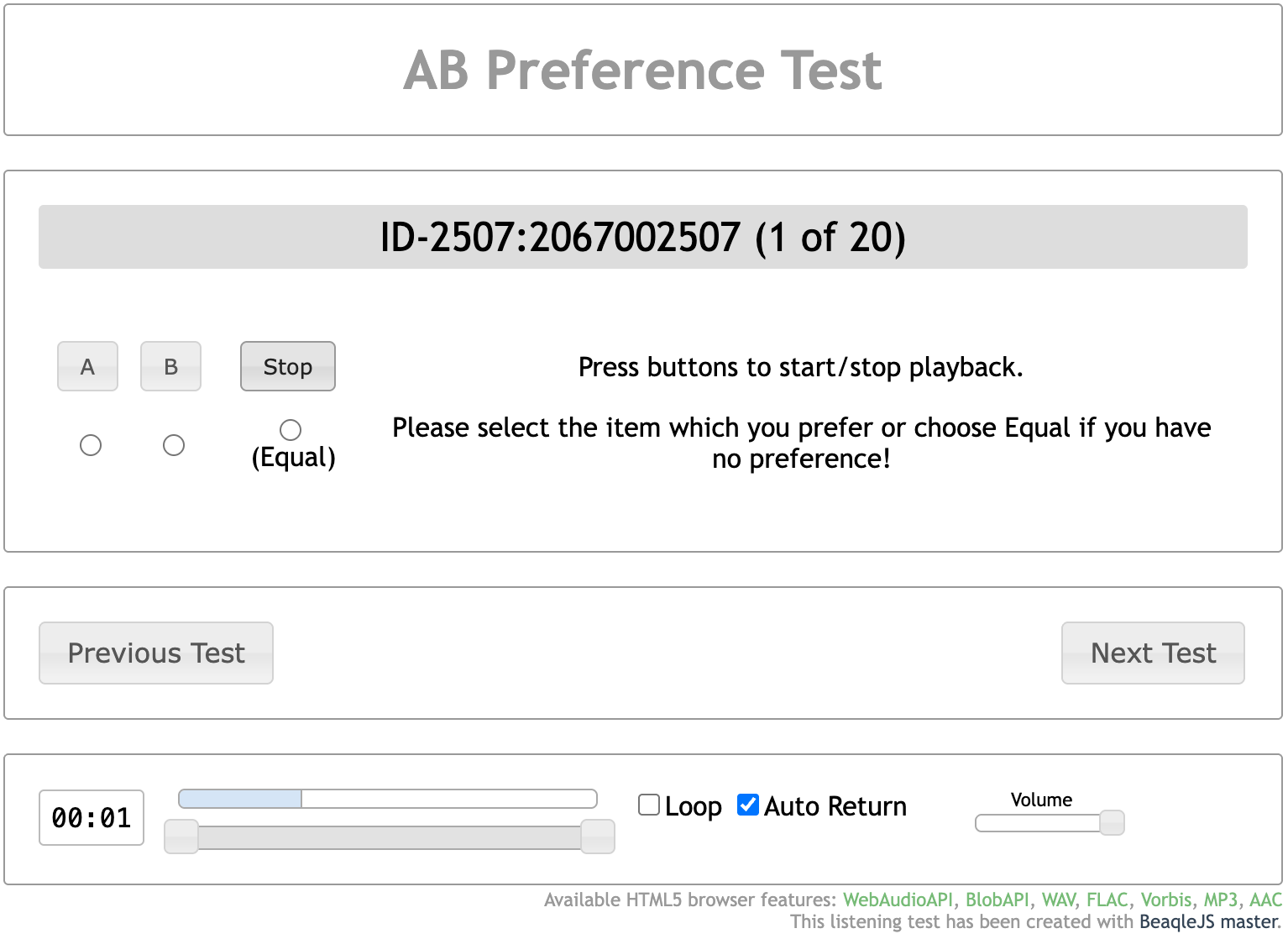}
    \caption{Screenshot of the AB preference test interface.}
    \label{fig:ab_test_screenshot}
\end{figure}

\end{document}